\documentclass[%
a4paper,							
11pt,								
bibliography=totoc,						
abstracton,					
]
{scrartcl}

\usepackage[a4paper,left=3.0cm,right=2.5cm, top=2.5cm, bottom=3.0cm]{geometry}
\usepackage[headsepline=.4pt,footsepline=.4pt,automark,autooneside=false,]{scrlayer-scrpage}
\clearpairofpagestyles
\automark[subsection]{section} 
\automark*[subsubsection]{subsection}
\ihead{\scriptsize\rightmark} 
\usepackage[ngerman, english]{babel}
\usepackage[T1]{fontenc}
\usepackage[utf8]{inputenc}
\usepackage{lmodern} 
\usepackage[onehalfspacing]{setspace} 

\usepackage{xspace}
\usepackage{calc}

\usepackage{amsmath}
\usepackage{amssymb}
\usepackage{amsthm,thmtools}

\usepackage{mathtools}
\usepackage{bbm}
\usepackage{bm}
\begingroup
    \makeatletter
    \@for\theoremstyle:=definition,remark,plain\do{%
        \expandafter\g@addto@macro\csname th@\theoremstyle\endcsname{%
            \addtolength\thm@preskip\parskip
            }%
        }
\endgroup

\usepackage{chngcntr}
\counterwithin*{equation}{section}

\theoremstyle{plain}
\newtheorem{theorem}{Theorem}[section]

\theoremstyle{definition}

\theoremstyle{remark}
\newtheorem{remark}[theorem]{Remark}
\theoremstyle{plain}

\theoremstyle{plain}

\theoremstyle{remark}

\usepackage{tabularx}
\usepackage{booktabs}
\usepackage{multirow}
\usepackage{multicol}
\usepackage{pdflscape}

\usepackage{diagbox}

\usepackage[flushleft]{paralist}

\usepackage{color}

\usepackage{graphicx}
\usepackage{tikz}
\usetikzlibrary{calc,spy,backgrounds,patterns,arrows}
\usepackage{todonotes}
\usepackage[final]{showkeys}

\usepackage{fancyvrb}

\usepackage{float}
\usepackage[section]{placeins}
\usepackage[format=plain]{caption}
\usepackage{csquotes} 

\usepackage{titleref}
\usepackage{hyperref}
\usepackage{cleveref}
\hypersetup{
    colorlinks,
		linkcolor={red},
    citecolor={blue},
    urlcolor={blue}
}

\renewcommand{\P}{\mathbb{P}}

\newcommand{\R}{\mathbb{R}}

\newcommand{\N}{\mathbb{N}}

\newcommand{\lieAlgebra}{\mathfrak{g}}
\newcommand{\lieBasis}{\mathcal{E}}
\newcommand{\lieGroup}{G}

\newcommand{\ad}{\mathrm{ad}}

\newcommand{\norm}[1]{\left\|#1\right\|}
\newcommand{\abs}[1]{\left|#1\right|}
\newcommand{\vectorize}{\mathrm{vec}}

\newcommand{\rFormat}[1]{\text{\textbf{#1}}\xspace}
\newcommand{\rHistorical}{R^{\mathrm{H}}}
\newcommand{\rGAN}{R^{\mathrm{GAN}}}
\newcommand{\rSDE}{R^{\mathrm{SDE}}}
\newcommand{\rCIR}{R^{\mathrm{CIR}}}
\newcommand{\rGEM}{R^{\mathrm{gEM}}}

\newcommand{\tabAngle}{45}
\newcommand{\tabTextWidth}{3cm}
\newcommand{\rotateTableHead}[1]{\rotatebox{\tabAngle}{\parbox[c]{\tabTextWidth}{\centering #1}}}
\newcolumntype{C}{X<{\centering}}

\newcommand{\CPU}{Intel(R) Core(TM) i7-8750H CPU @ 2.20\,GHz\xspace}
\newcommand{\RAM}{2x32\,GB (Dual Channel) Samsung SODIMM DDR4 RAM @ 2667 MHz\xspace}
\newcommand{\GPU}{NVIDIA GeForce RTX 2070 with Max-Q Design (8\,GB GDDR6 RAM)\xspace}
\newcommand{\OS}{Windows 10 Pro\xspace}

\SaveVerb{python}=(Intel-)Python 3.9=
\newcommand{\python}{\protect\UseVerb{python}\xspace}
\SaveVerb{tensorflow}=Tensorflow 2.8.0=
\newcommand{\tensorflow}{\protect\UseVerb{tensorflow}\xspace}
\SaveVerb{matlab}=Matlab 2022a=
\newcommand{\matlab}{\protect\UseVerb{matlab}\xspace}
\SaveVerb{ga}=ga=

\SaveVerb{fmincon}=fmincon=

\SaveVerb{fitdist}=fitdist=

\SaveVerb{lsqnonlin}=lsqnonlin=
\newcommand{\lsqnonlin}{\protect\UseVerb{lsqnonlin}\xspace}
\newcommand{\matlabGOtoolbox}{(Global) Optimization Toolbox\xspace}

\newcommand{\suffix}{-eps-converted-to.pdf}
\title{A novel approach to rating transition modelling via Machine Learning and SDEs on Lie groups}
\author{Kevin Kamm\thanks{Dipartimento di Matematica, Universit\`a di Bologna, Bologna, Italy.
\textbf{e-mail}: kevin.kamm@unibo.it}
\and Michelle Muniz\thanks{Institute of Mathematical Modelling, Analysis and Computational Mathematics (IMACM), 
Chair of Applied Mathematics and Numerical Analysis, Bergische Universität Wuppertal, Wuppertal, Germany.
\textbf{e-mail}: muniz@uni-wuppertal.de}
}

\begin{document}
\thispagestyle{empty}\pagenumbering{roman}
\maketitle
\renewcommand{\thefootnote}{\Roman{footnote}}
\renewcommand{\thefootnote}{\arabic{footnote}}
\begin{abstract}
In this paper, we introduce a novel methodology to model rating transitions with a stochastic process.
To introduce stochastic processes, whose values are valid rating matrices, we noticed the geometric properties of stochastic matrices and its link to matrix Lie groups. We give a gentle introduction to this topic and demonstrate how Itô-SDEs in $\R$ will generate the desired model for rating transitions.

To calibrate the rating model to historical data, we use a Deep-Neural-Network (DNN) called TimeGAN to learn
the features of a time series of historical rating matrices. Then, we use this DNN to generate synthetic rating transition matrices. Afterwards, we fit the moments of the generated rating matrices and the rating process at specific time points, which results in a good fit.

After calibration, we discuss the quality of the calibrated rating transition process
by examining some properties that a time series of rating matrices should satisfy, and we will see that this geometric approach works very well.
\end{abstract}
\noindent
\textbf{Acknowledgements:}
This project has received funding from the European Union’s Horizon 2020 research and innovation programme
under the Marie Sklodowska-Curie grant agreement No 813261 and is part of the ABC-EU-XVA
project.\\\noindent
\textbf{Code availability:}
The code and data sets to produce the numerical experiments are available at
\url{https://github.com/kevinkamm/RatingML}.
\newpage
\pagestyle{scrheadings}\ihead{\scriptsize\rightmark}\pagenumbering{arabic}
\section{Introduction}\label{sec:introduction}
In this paper, we model rating transition matrices with a stochastic process using historical data published by rating agencies such as S\&P, Moody's or Fitch for the calibration. 

This is done in two steps.
First, we show how a Deep-Neural-Network (DNN) known as TimeGAN (cf. \cite{Yoon2019})
can be utilized to learn the distribution of the historical rating transitions. 
In a second step, we match the moments of the model and the synthetic data generated by the DNN. For the stochastic model itself, we will demonstrate how basic matrix Lie group theory can be helpful to define Itô-processes in $\R$ to model the rating transitions.

A rating is an indicator of the creditworthiness of an entity. A high rating associates less risk to an entity to not fulfill its financial obligations and a low rating a high risk.
Ratings are usually denoted by letters
$\rFormat{A}$, $\rFormat{B}$, \dots, $\rFormat{D}$, where $\rFormat{A}$ denotes the best rating and
$\rFormat{D}$ denotes the worst rating. The rating $\rFormat{D}$ is special. It means that an entity has defaulted, i.e.\ it can not fulfill its financial obligation towards a contracting party. In this paper, we use the terms default and bankruptcy of an entity synonymous, implying that a defaulted company cannot recover from this state.

For most applications, it is important to model the rating changes of an individual entity or an entire sector on a continuous time scale. 
This can be done in two different ways. On the one hand, one can define a process $X_t$, which tells us at each time and trajectory the current rating of a company. The natural state-space of these processes is therefore discrete and the time axis is continuous.
On the other hand, one can model the transition probabilities $R_t$ of a sector at each point in time and derive a rating process using these transition probabilities. The state-space of this type of model is then a matrix whose entries are the probability of transitioning from one rating to another starting at an initial time $t_0$ (usually today) till a future time $t$. An example of such a $t-t_0$ rating matrix is given in \Cref{tab:ratingMatrix1}.
\begin{table}[h]
    \centering
    \begin{tabular}{|c|*{4}{c}|} 
        \hline
        \diagbox[]{From}{To}  & \rFormat{A} & \rFormat{B} & \rFormat{C} & \rFormat{D}\\
        \hline
        \rFormat{A} & 0.9395 & 0.0566 & 0.0037 & 2.7804e-04\\ 
        \rFormat{B} & 0.0092 & 0.9680 & 0.0211 & 0.0017\\ 
        \rFormat{C} & 6.2064e-04 & 0.0440 & 0.8154 & 0.1400\\ 
        \rFormat{D} & 0 & 0 & 0 & 1\\
        \hline
    \end{tabular}
    \caption{Example of a one year rating transition matrix.}
    \label{tab:ratingMatrix1}
\end{table}
We can see that the individual rows sum up to one, meaning that all rows are valid probability distributions. These type of matrices are called stochastic for this reason. The last row corresponds to our idealized assumption that a defaulted entity cannot recover, i.e.\ the default state is absorbing.
Rating agencies publish these type of matrices usually once a year for a few time frames. Short-term rating matrices are usually published with time frames of $1,3,6,12$ months and long-term rating matrices
with time frames of $1,2,3,5,10$ years.
We see a lot of uncertainty in the historical data published by the agencies increasing with larger time frames. Therefore, it stands to reason to desire stochastic models for the rating transitions.

Thus, we would like to model the evolution of rating transition matrices as seen from today with a stochastic process in continuous time. We will focus in this paper on short-term rating matrices.

\subsection{Review of the literature and comparison}\label{sec:review}
We recognize two different approaches to rating modelling in the literature which is described in
\cite[p.~76 Section 4.12.1 Standing Assumptions]{Bielecki2003} in more details.
On the one hand, one can model ratings in a HJM-framework, independently proposed by \cite{Bielecki2000}
and \cite{Schonbucher2003}. On the other hand, there are intensity-based models, introduced by the pioneering work of \cite{Jarrow1997}. As this paper can also be viewed as an intensity approach let us explain this in more details alongside a short illustration in \Cref{fig:litVsOur1}.
\newlength{\introX}\setlength{\introX}{6cm}
\newlength{\introY}\setlength{\introY}{2cm}
\tikzset{introNode/.style={black,rectangle, thick, text width = .5\introX, align=center},
         introArrow/.style={thick}}

\begin{figure}[h]
\centering
\begin{tikzpicture}
\node[introNode] (center) at (0,0) {Rating Transition Matrices};
\node[introNode] (litR) at ($(-\introX,\introY)+(center)$) {$X_t$ CTMC};
\node[introNode] (litU) at ($(-\introX,0)+(center)$) {$U_{t}$ Transition operator};
\node[introNode] (litA) at ($(-\introX,-\introY)+(center)$) {$A$ generator};
\node[introNode] (ourR) at ($(\introX,\introY)+(center)$) {$X_t$ rating process};
\node[introNode] (ourU) at ($(\introX,0)+(center)$) {$R_{t}$ process in Lie Group};
\node[introNode] (ourA) at ($(\introX,-\introY)+(center)$) {$L_t$ SDE in Lie Algebra};

\draw[->,introArrow] (litR) -- (litU);
\draw[->,introArrow] (litU) -- (litA);

\draw[->,introArrow] (ourU) -- (ourR);
\draw[->,introArrow] (ourA) -- (ourU);

\draw[<->,introArrow] (litU) -- node[midway,above]{Calibration} (center);
\draw[<->,introArrow] (center) -- node[midway,above]{Calibration} (ourU);
\end{tikzpicture}  
\caption{Illustration how our approach compares to the literature.}
\label{fig:litVsOur1}
\end{figure}
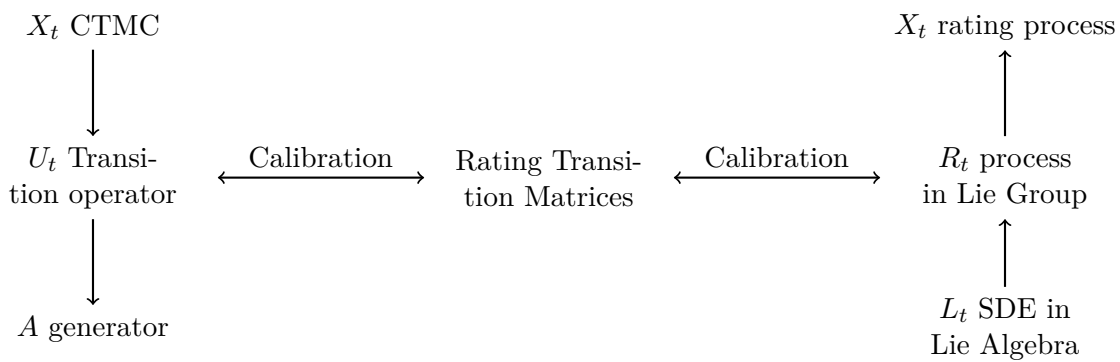
In the intensity approach (left-hand side in \Cref{fig:litVsOur1}), usually the rating process $X_t$ is modelled by a continuous-time Markov chain (CTMC). This seems quite natural, because its state space is discrete. Another feature of this approach is that due to the Markovianity one can describe a CTMC fully by its transition operators $U_t$. Transition operators tell us for a given initial time and state the probability to transition to another state at a later time. So exactly, what rating transition matrices describe. Assuming time-homogeneity of the CTMC, it is easy to derive a so-called generator $A$ of the transition operator, which gives a full characterization of the CTMC. This leads to an analytical and numerical tractable model.

However, in this setting the transition operators and generators are deterministic and in the special case of homogeneous CTMCs (the most common assumption in the literature), the generator is constant. 
While this makes it possible to calibrate the model directly to the published rating matrices, it limits the possibility for modelling time-dependent features or uncertainty. 

In this paper, we want to model the rating transitions with a stochastic process (right-hand side in \Cref{fig:litVsOur1}) and noticed that generators of CTMCs are actually elements in a suitable subspace of the Lie algebra of stochastic matrices. This allows us to formulate Itô-SDEs taking values in $\R_{\geq 0}$ and apply a basis transformation to the desired Lie algebra leading to a process $L_t$. The exponential map, i.e. the matrix exponential, maps the model in the Lie algebra to the proper Lie group of stochastic matrices resulting in a stochastic model $R_t$.
For the calibration, we need to study the distribution of the time series of historical rating matrices, for which we use a TimeGAN.

To the best of our knowledge, this is the first paper which is modelling rating transitions starting from an SDE in a appropriate subspace of the Lie algebra of stochastic matrices. Additionally, the application of a Deep-Neural-Network (DNN) to learn the distribution of historical rating transition matrices seems entirely novel in this community. Also we believe that this is an exciting approach with many possibilities for future research from both a theoretical point of view and modelling point of view.

The paper is structured as follows: In \Cref{sec:fakeRM} we will train a DNN learning a time series of $1,3,6,12$ rating matrices. The section is divided into two parts. In \Cref{sec:AalenJohansen} we explain how to compute rating matrices from historical data making certain that all rows sum up to one. This is followed in \Cref{sec:timeGAN} by a description how the training data is built and how the TimeGAN DNN works.
In \Cref{sec:SDELieGroup} we give a gentle introduction to matrix Lie groups and notice that the stochastic matrices form a are a subgroup of matrix Lie group. We show two different ways how to utilize this framework to model rating transition matrices by a stochastic process.
Afterwards, we do some numerical experiments in \Cref{sec:numerics} and define desirable properties of short-term rating matrices in \Cref{sec:errors}. The first step is to calibrate the rating process to the distributions learned by the DNN, which is subject of \Cref{sec:calibration}. Then, in \Cref{sec:directExp} and \Cref{sec:gEM} we perform one test for each of the two methods proposed in
\Cref{sec:SDELieGroup} and assess their quality.
Last but not least, we conclude the paper in \Cref{sec:Conclusion} and discuss possibilities for future research.

\section{Generating rating transition matrices}\label{sec:fakeRM}
In this section, we will explain how to generate synthetic rating transition matrices from historical data.
The section is structured as follows. First of all, we will discuss in \Cref{sec:AalenJohansen} what rating matrices are, what kind of historical data we have and how to compute them. Afterwards, we will give a brief introduction to the relevant Deep-Neural-Network (DNN) architectures, which are necessary for the TimeGAN in \Cref{sec:timeGAN}.
\subsection{Historical data and Aalen-Johansen estimator}\label{sec:AalenJohansen}
Ratings are an ordered set of indicators for creditworthiness of an entity. The best rating is usually denoted by the letter \rFormat{AAA} or simply \rFormat{A}. If an entity is insolvent, meaning that it cannot fulfill its financial obligations, we say that this entity has defaulted. In this paper, we will not distinguish between the default and bankruptcy of an entity, which translates to the fact that once an entity has defaulted, it cannot recover from it. In the mindset of ratings, a default can be viewed as the worst possible rating usually denoted by \rFormat{D}.

To keep this presentation as simple as possible in this paper, we consider only four different ratings: \rFormat{A}, \rFormat{B}, \rFormat{C}, \rFormat{D} ordered from best to worst rating and identify them by integers $\left\{1,2,\dots,K\right\}$, whenever it is more convenient. But it is straightforward to use more ratings.
\paragraph*{Methodology.}
Rating agencies, such as S\&P, Moody's and Fitch are required by \enquote{Rule 17g-7 of the Securities Exchange Act of 1934}\footnote{Please visit \url{https://www.sec.gov/structureddata/rocr-publication-guide.html} for more details. Last accessed: 19.05.2022 12:23.} to publish the history of rating changes for some entities.
This data can be downloaded from their respective websites and consists of rating histories of individual entities in different sectors, e.g. financial institutes and corporate. We will use the data set from S\&P with focus on the corporate sector. The data is structured like follows: for each entity it consists of a list of time stamps when a rating was changed or confirmed. Therefore, we can extract the historical ratings for each individual company for each day. 

After extracting these rating trajectories, we apply the so-called Aalen-Johansen estimator (cf. \cite{Lando2002}) to the processed data to compute the rating transition matrices with a given time span. For example, we can set our initial time to the first of January of a specific year and compute the rating transitions over one year to get an average rating transition matrix of one year in the corporate sector.

Let us explain this in more details. The Aalen-Johansen estimator is a non-parametric estimator of the transition probabilities of a time-inhomogeneous continuous-time Markov chain (ICTMC) and we will assume that the historical rating transition data can be modelled by an ICTMC. The rating transition probabilities starting at time $s$ up to time $t$ are then estimated by
\begin{align*}
    P\left(s,t\right) \coloneqq 
    \prod_{k=1}^{m}{
        \left(
            I + \Delta A\left(T_k\right)
        \right)
    },
\end{align*}
where $T_k$ is the jump time in the interval $\left[s,t \right]$ and $m\in\N$ is the number of jumps, 
as well as the estimated generator
\begin{align*}
    \Delta A\left(T_k\right) \coloneqq
    \left(
        \begin{array}[c]{*{5}{c}}
             -\frac{\Delta N_{1}\left(T_k\right)}{Y_{1}\left(T_k\right)}&
                \frac{\Delta N_{12}\left(T_k\right)}{Y_{1}\left(T_k\right)}&
                \frac{\Delta N_{13}\left(T_k\right)}{Y_{1}\left(T_k\right)}&
                \cdots &
                \frac{\Delta N_{1K}\left(T_k\right)}{Y_{1}\left(T_k\right)}\\
            \frac{\Delta N_{21}\left(T_k\right)}{Y_{2}\left(T_k\right)}&
                -\frac{\Delta N_{2}\left(T_k\right)}{Y_{2}\left(T_k\right)}&
                \frac{\Delta N_{23}\left(T_k\right)}{Y_{2}\left(T_k\right)}&
                \cdots &
                \frac{\Delta N_{2K}\left(T_k\right)}{Y_{2}\left(T_k\right)}\\
             \vdots &
             \vdots &
             \ddots &
             \cdots &
             \vdots\\ 
             \frac{\Delta N_{K-1,1}\left(T_k\right)}{Y_{K-1}\left(T_k\right)}&
                \frac{\Delta N_{K-1,2}\left(T_k\right)}{Y_{K-1}\left(T_k\right)}&
                \cdots&
                -\frac{\Delta N_{K-1}\left(T_k\right)}{Y_{K-1}\left(T_k\right)}&
                \frac{\Delta N_{K-1,K}\left(T_k\right)}{Y_{K-1}\left(T_k\right)}\\
            0 & 0 & \cdots & \cdots & 0
        \end{array}
    \right).
\end{align*}
The jump process $\Delta N_{ij}\left(T_k\right)$ denotes the number of transitions from rating $i$
to rating $j$ at time $T_k$ and $\Delta N_{i}\left(T_k\right)$ counts the total number of transitions away from rating $i$ at time $T_k$.
The jump process $Y_i\left(T_k\right)$ denotes the number of entities with rating $i$ right before time $T_k$. 
The last row is zero, because we assume an absorbing default rating.
So each time a rating changes in the underlying data, the estimated generator is updated accordingly.

For a more detailed explanation with examples we refer to \cite[pp.~9\,ff.]{Lando2002}.
\paragraph*{Advantages and limitations.} 
To discuss the advantages of using the Aalen-Johansen estimator, we need to briefly discuss a huge problem of the rating data. Entities have the right at any point in time to not being rated anymore for whatever reason. This is a huge issue, because suppose you would want to calculate the rating transition probabilities naively by setting a time frame, denote how many companies are in which rating initially and then look where they end up at the end of the time frame. If a company decides to withdraw from being rated in this time window, one has at the end a rating matrix with rows that do not sum up to one, i.e.\ an invalid probability distribution. The Aalen-Johansen estimator overcomes this problem naturally, by updating after each rating change. Therefore, this method guarantees that rows sum up to one, which will be important later on.

However, we found that our results differ from the rating matrices which are published by the agencies and confirmed with S\&P that they also use unpublished sensitive rating data and remove correlation structures from data, for which additional knowledge of the entities and their relation towards each other is necessary.

Therefore, the results presented in this paper serve as an illustration how this methodology can be applied but the underlying data needs some work for an implementation in practice. 

\subsection{TimeGAN}\label{sec:timeGAN}
In this section, we show how one can use a generative adversarial network (GAN) for time series data to 
obtain fake rating transition matrices from paths of a Brownian motion. In particular, we chose 
a network called TimeGAN by \cite{Yoon2019} to learn the rating distributions from the historical data.

The TimeGan is supposed to learn a function 
\begin{align*}
    f(t_k,W_{t_k}\left(\omega\right)) = R_{t_k}\left(\omega\right)
\end{align*}
mimicking the historical rating matrices $\rHistorical_{t_k}$ for $k=1,\dots,n$, $n\in \N$. After the learning phase,
we can use a path of the Brownian motion $W$ to generate fake rating matrices at the points in time 
$t_k$.

\paragraph*{Training data.} We use the technique described in the previous paragraph to compute rating matrices with time spans of $1,3,6,12$ months starting in 2011 till the end of 2019. For the one month rating matrices, we start at each month in a year and compute the transition probabilities with the Aalen-Johansen estimator till the next month. For the three month rating matrices we proceed similar but starting every three months and so on, such that data is not used twice for the rating matrices with respective time spans. After computing all these matrices we end up with 108 matrices for one month, 36 for three months, 
18 for six months and nine for one year. 
After that, we build a set of time series data by considering all the permutations of the rating matrices leading to a data set of roughly 630000 different time sequences of rating matrices.

We are aware that this approach might raise some eyebrows but rating data is scarce and it is not unusual to assume independence of the rating events which justifies this approach. We will discuss the impact of this choice in \Cref{sec:numerics} further, while studying properties of rating matrices.

\begin{remark}\label{rem:dataset}
One can alternatively use the rating matrices which are published by the rating agencies from e.g.\ the last 10 years. However, these are usually only available for long term rating matrices, i.e.\ 1 up to 10 years. Another problem with this data set is that rating agencies use the so-called cohort method to compute the matrices, i.e.\ they have the imperfections due to entities who do not want to be rated anymore. So one idea could be to repair them with an heuristic method and build up a training data set by again considering the permutations of the time series.
\end{remark}

The TimeGAN combines an autoencoder with a generative adversarial network using recurrent neural networks linked by a supervising network. We would like to give a short intuition how these networks work together in our case and refer the reader to \cite{Yoon2019} for the details.

\begin{figure}[h]
    \centering
    \includegraphics[width=.75\columnwidth]{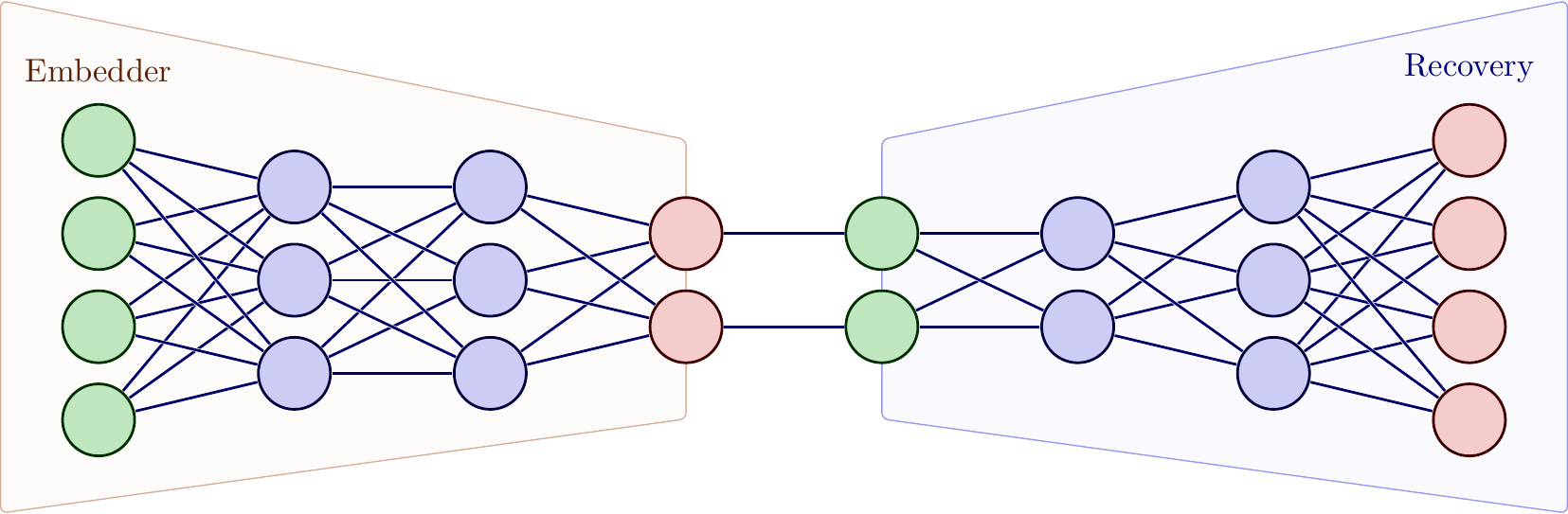}
    \caption{Illustration of a VAE network.}
    \label{fig:VAE1}
\end{figure}
\begin{figure}[h]
    \centering
    \includegraphics[width=.75\columnwidth]{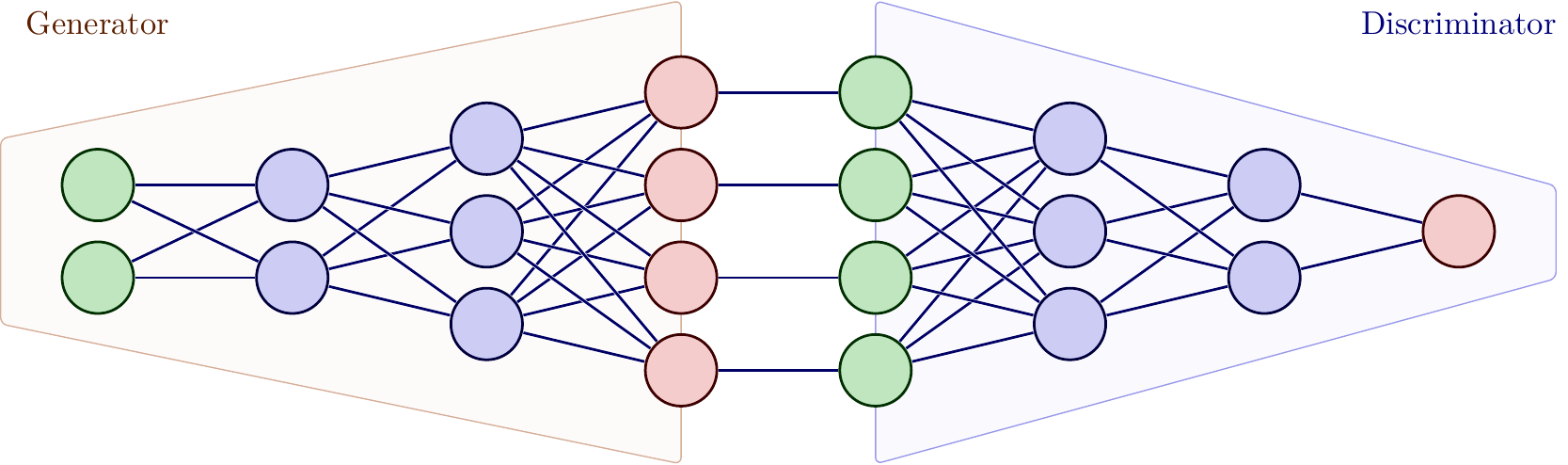}
    \caption{Illustration of a GAN network.}
    \label{fig:GAN1}
\end{figure}

\paragraph*{Autoencoder.}
For a detailed treatment of Variational Autoencoders (VAE) we refer the reader to \cite{Kingma2019}.

The principle network architecture in an application without time series data is illustrated in
\Cref{fig:VAE1}. There are two different networks linked to each other, one called \emph{embedder} or \emph{encoder} and the other one called \emph{recovery} or \emph{decoder}. The idea is to introduce a bottleneck between these networks. This forces the network to learn principle components of the data and helps with denoising as well as dimensionality reduction. For the training phase, the data is first embedded, recovered and afterwards compared to the original data to minimise the difference of both.
After the training phase the recovery network can be used to generate rating matrices from their embedded features. We will see how the generator network of generative adversarial network can be used to generate fake features in the next two paragraphs.

\paragraph*{Generative Adversarial Network.}
For a detailed treatment of Generative Adversarial Networks (GAN) we refer the reader to \cite{Goodfellow2017}.

The principle network architecture in an application without time series data is illustrated in
\Cref{fig:GAN1}. There are two different networks linked to each other, one called \emph{generator} and the other one called \emph{discriminator}. The idea is to play these networks against each other. The generator network has a few random numbers as input and outputs fake data. The discriminator network will get the fake data from the generator as an input, as well as the real data. Then it is learning to distinguish between fake and real data by outputting a probability of the data being real. Since we know which of the input data is fake and which is real we can optimize the prediction of the discriminator network. The generator on the other hand is learning how to fool the discriminator, i.e.\ making it believe that the fake data point was real.
After the learning phase and when the discriminator is not very confident anymore in distinguishing between fake and real, the generator network can be used to produce synthetic data.

\paragraph*{Supervisor.}
The supervisor network does not have a special network architecture and it is placed in-between the embedder and recovery network, as well as between the generator and the discriminator network to establish a link between them. This makes it also possible in the training of the entire network to compare the supervised networks to the unsupervised networks. Another implication of this approach is that the generator network of the GAN is not generating the rating matrices directly but the features of the rating matrices.
As aforementioned, combining the trained generator with the trained recovery network will enable us to generate synthetic rating matrices.

\paragraph*{Recurrent networks.}
For a detailed treatment of Recurrent Neural Networks (RNN) and a comparison of Long-Short-Term-Memory (LSTM) to Gated-Recurrent-Units (GRU) we refer the reader to \cite{Chung2014}.

So far, we have discussed how the supervised VAE and GAN can be used together at a single point in time to generate synthetic rating matrices. RNNs enable us to use time series data and all the aforementioned networks are augmented with GRUs in our implementation to take the time series of rating matrices into account.
GRUs consist of two different gates. One is called the \emph{update gate} and the other one is called \emph{forget gate}. The update gate decides how much of the new temporal information is added to the time sequence. The forget gate has the possibility to forget the previous times in the time sequence, making the current point in time independent of the past.

\paragraph*{Hyperparameters and network architecture.} It is not the purpose of this paper to \enquote{over-optimize} the procedure, since it is a first step using these modern techniques for rating transitions. Additionally, for its next use case of rating triggers, an additional source of market data will be available and the current architecture might need some adjustments. We leave it up to the reader to change the hyperparameters and network architectures, because we are satisfied with the performance of the current setting, which is discussed in \Cref{sec:errors} in greater detail. We chose the following settings for our experiments.
\begin{compactenum}
    \item We used 40 epochs in total and noticed that 10 epochs take roughly 1 hour in the training step.
    \item We found that a batch size of 128 was a good middle-ground between speed and realistic rating matrices.
    \item For the embedder we used three GRU layers. The first and last with 3 units and the second one with 2 units. The output dense layer has 4 units and a sigmoid activation function.
    \item For the recovery we used three GRU layers. The first and last with 3 units and the second one with 2 units. The output dense layer has $K^2=16$ units and a sigmoid activation function.
    \item For the supervisor we used two GRU layers, each with 4 units. The output dense layer has 4 units and a sigmoid activation function.
    \item For the generator we used three GRU layers, each with 4 units. The output dense layer has 4 units and a sigmoid activation function. As an input we take the values of a Brownian path at
    $t=1,3,6,12$ months.
    \item For the discriminator we used three GRU layers, each with 4 units. The output dense layer has a single unit and a sigmoid activation function.
    \item All optimizers were Adam (cf. \cite{Kingma2014}) with the standard learning rate $1e-4$.
\end{compactenum}
As aforementioned, for the training of the network we refer the reader to
\cite{Yoon2019} and note that we used the standard loss functions indicated in this paper.

\section{SDEs on the Lie Group of stochastic matrices}\label{sec:SDELieGroup}
In this section, we show how an SDE can help to interpolate the generated rating matrices in time.
This is a desirable feature for several applications, because it gives access to rating matrices of any time span or can help to forecast transition matrices with larger time spans.

To guarantee that our SDE will produce stochastic matrices, we noticed that this is a special kind of geometry and the proper tools are readily available in the matrix Lie group literature. We will recall all the necessary results first.

\begin{figure}
    \centering
    \begin{tikzpicture}
        \path[pattern=north west lines, pattern color=green, opacity=.25] 
            (-11,0) rectangle (-8,1.5) 
            node[pos=.5, opacity=1] (rPos) {$\R^{(K-1)^2}_{\geq 0}$};
        \draw[dashed] (-11,-1.5) rectangle  (-8,1.5);
        \node[] at (-9.5,0) {$\R^{(K-1)^2}$};
        
        \path[pattern=north west lines, pattern color=blue, opacity=.25] 
            (-6,0) rectangle (-3,2) 
            node[pos=.5, opacity=1] (algebraPos) {$\lieAlgebra_{\geq 0}$};
        \draw[dashed] (-6,-2) rectangle  (-3,2);
        \node[] at (-4.5,0) {$\lieAlgebra$};
        
        \draw[dashed] (.5,0) ellipse (1.5cm and 2cm)
             node[pattern=north west lines, pattern color=red,circle,opacity=.4] 
             (lieGroupPos) {
             \tikz{\node[rectangle,opacity=1]{$\lieGroup_{\geq 0}$};}};
        \node[] at (.5,-1.25) {$\lieGroup$};
        
        \draw[->,>=stealth',auto] (algebraPos) to[bend left] node[sloped,below] {$\exp$} (lieGroupPos) ;
        \draw[<->,>=stealth',auto] (rPos) -- node[sloped,anchor=center, align=center,text width=3cm]{Coordinates $L^i$\\Basis $\lieBasis_i$} (algebraPos);
        
    \end{tikzpicture}
    \caption{Illustration of the relationship between $\R^{(K-1)^2}_{\geq 0}$, $\lieAlgebra_{\geq 0}$ and $\lieGroup_{\geq 0}$.}
    \label{fig:lie1}
\end{figure}
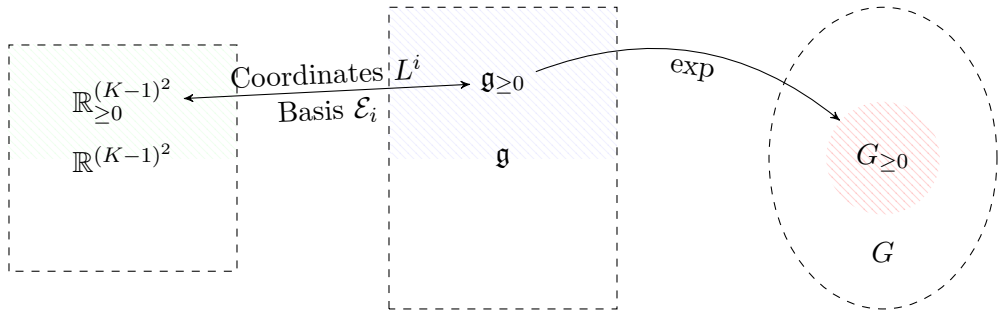

We consider the group $\lieGroup=\{R\in\mathrm{GL}(K): R\bm{1}=\bm{1}\}$, $\bm{1}=[1,\dots,1]^\top\in \R^K$, which is a matrix Lie group according to \cite{Coletti2020}, i.e.\ a subgroup of GL($K$) which is a differentiable manifold and for which the product is a differentiable mapping $\lieGroup\times \lieGroup\to \lieGroup$. 
The tangent space at the identity of a Lie group is called the Lie algebra and is in this case given by $\lieAlgebra=T_I \lieGroup=\{L\in \mathrm{GL}(K):L\bm{1}=\bm{0}\}$.
The Lie algebra $\lieAlgebra$ is a vector space with $\mathrm{dim}(\lieAlgebra)=K(K-1)$ since  basis matrices for $\lieAlgebra$ can be formulated as $E_{ij}-E_{ii}$ for $i,j=1,\dots,K$ with $i\neq j$, where $E_{ij}$ are elementary matrices.
This makes the Lie algebra $\lieAlgebra$ together with the matrix commutator, $[\cdot,\cdot]\colon\lieAlgebra\times\lieAlgebra\to\lieAlgebra$, $[L_1,L_2]=L_1L_2-L_2L_1$, isomorphic to $\mathbb{R}^{K(K-1)}$.
The matrix exponential $\exp\colon\lieAlgebra\to \lieGroup$, $\exp(L)=\sum_{k=0}^{\infty}L^k/k!$, maps elements from the Lie algebra to the Lie group and is a local diffeomorphism in a neighbourhood of $L=0$. 
The directional derivative of the matrix exponential along an arbitrary matrix $H\in\lieAlgebra$ is given by
\begin{equation*}
    \left(\frac{d}{dL}\exp(L)\right)H = \exp(L)d\exp_{-L}(H) \quad \text{with }\; d\exp_{-L}(H) = \sum_{k=0}^{\infty}\frac{1}{(k+1)!}\ad_{-L}^k(H).
\end{equation*}
where $\ad_{L}\colon\lieAlgebra\to\lieAlgebra$, $\ad_{L}(H)=[L,H]$ denotes the adjoint operator, which is used iteratively,
\begin{equation*}
    \ad_{L}^0(H)=H, \quad \ad_{L}^k(H)=\ad_{L}\big(\ad_{L}^{k-1}(H)\big)=\lbrack L,\ad_{L}^{k-1}(H)\rbrack
\end{equation*}
for $k\geq1$.
For more details on Lie groups and Lie algebras we refer the interested reader to \cite{Hall03}.

Consider the following SDE in the Lie algebra $\lieAlgebra$
\begin{align}\label{eq:SDELieAlgebra}
	dL_t = A(t,L_t) dt + B(t,L_t) dS_t,\quad L_0=0,
\end{align}
where $A,B\in \lieAlgebra$ and $S_t$ is a one-dimensional general semimartingale.
Applying a numerical scheme, e.g. the Euler-Maruyama scheme, to get an approximation $L_{t_{k+1}}$ of \eqref{eq:SDELieAlgebra} after one time step and computing $R_{t_{k+1}}=R_{t_k}\exp(L_{t_{k+1}})$  would result in a numerical method for solving
\begin{equation}\label{eq:SDELieGroup}
    dR = \left(R\, d\exp_{-L}(A) + \frac{1}{2}\Big(\frac{d}{dL}R\, d\exp_{-L}(B)\Big)B\right) dt + R\,d\exp_{-L}(B) \,dW_t, \quad R_0=I,
\end{equation}
which can be easily verified by applying Itô's lemma to $R_t=R_0\exp(L_t)\in \lieGroup$ in the case $S_t=W_t$ a Brownian motion (as done e.g. in \cite{KPP2021}).
As this approach preserves the geometry of the Lie group $\lieGroup$ opposed to applying the Euler-Maruyama scheme directly to \eqref{eq:SDELieGroup}, this method was called the \textit{geometric Euler-Maruyama} scheme in \cite{MaSo18}. Higher order schemes based on this approach can be found in \cite{Muniz2022}.

Since we are interested in stochastic matrices that are elements of $\lieGroup_{\geq 0}\coloneqq \{R\in \lieGroup: R_{ij}\in [0,1], i,j=1,\dots,K\}$ we now consider a subset of the Lie algebra $\lieAlgebra$, namely $\lieAlgebra_{\geq0}\coloneqq\{L\in\lieAlgebra:L_{ij}\geq0,i\neq j, L_{ii}\leq 0, L_{Kj}=0, i,j=1,\dots,K\}$. 
Note that additional to the usual properties of generator matrices we choose the last line of matrices $L\in\lieAlgebra_{\geq 0}$ to be zero because applying the matrix exponential $\exp$ to these matrices will generate matrices that have the last unit vector in the last line. 
This choice is in accordance with our assumption that the default state is absorbing.
With this assumption the dimension of $\lieAlgebra_{\geq 0}$ is now $\mathrm{dim}(\lieAlgebra_{\geq 0})=(K-1)^2$ because as before basis matrices can be denoted by $E_{ij}-E_{ii}$ but for $i=1,\dots,K-1$, $j=1,\dots,K$ and $i\neq j$. 
Similarly to before, there exists an isomorphism between $\mathfrak{g}_{\geq0}$ and $\mathbb{R}^{(K-1)^2}$, which is illustrated on the left-hand side in \Cref{fig:lie1}.
We will denote the basis for $\lieAlgebra_{\geq 0}$ by $\lieBasis_i$, $i=1,\dots,(K-1)^2$.
The fact that for any $L\in \lieAlgebra_{\geq 0}$ we have $\exp\left(L\right)\in \lieGroup_{\geq 0}$
is well-known and a proof can be found in \cite[pp.~86\,ff. Chapter 4.2.5: Solving Kolmogorov's Equation]{Stroock2005}.

\paragraph*{Direct exponential mapping.}
For the interpolation of the generated rating matrices we consider the SDE \eqref{eq:SDELieAlgebra} again and discuss some conditions for the solution $L_t$ to be evolving in $\mathfrak{g}_{\geq 0}$ such that $\exp(L_t)\in G_{\geq0}$.
Therefore, we make the assumption that the equation is decoupled in the following sense:
\begin{align}
    \begin{aligned}[c]
        dL_t &= A(t,L_t) dt + B(t,L_t) dS_t
             =\sum_{i=1}^{(K-1)^2}{
                \left(
                    \alpha_i(t,L^i_t) dt + \beta_i(t,L^i_t) dS_t
                \right) \lieBasis_i
             },
    \end{aligned}
    \label{eq:SDELie}
\end{align}
where $\lieBasis_i$ denotes the basis vectors of $\lieAlgebra_{\geq 0}$. If the solution $L_t^i$ of $dL_t^i = \alpha_i(t,L^i_t) dt + \beta_i(t,L^i_t) dS_t$ is $\P$-almost surely positive for all $t\geq 0$ and for all $i$ then $L_t \in \lieAlgebra_{\geq 0}$ and $\rSDE_t \coloneqq \exp\left(L_t\right) \in \lieGroup_{\geq 0}$.

Let us show two examples:
\begin{compactenum}
    \item Let $\alpha_i(t,x) \equiv a_i \in \R_{\geq 0}$, $\beta_i(t,x) \equiv b_i \in \R_{\geq 0}$:
        In this case, $L_t^i= a_i t + b_i S_t$ and the condition $L_t^i \geq 0$ leads to
        $a_i t + b_i S_t \geq 0$ for all $t$ $\P$-almost surely. Further assuming $S_t \geq 0$ would be one example.
    \item $L_t^i$ are CIR-processes, i.e. $S_t = W_t$ and
        $dL_t^i = a_i \left(b_i - L_t^i \right) dt + \sigma_i \sqrt{L_t^i} dW_t$.

\end{compactenum}

For this simple approach there is a price to pay, namely $\rSDE_t$ cannot be viewed as an evolution system of a Markovian rating process, since the Chapman-Kolmogorov equation is not necessarily satisfied.  Or in other words, the associated rating process will not be memoryless and it is difficult to sample it.

\paragraph*{Geometric Euler-Maruyama.}
In order to preserve the Chapman-Kolmogorov equation one could use the aforementioned geometric Euler-Maruyama scheme and define $\rSDE_t=R_0\exp(L_t)$. 
However, to ensure that $\rSDE_t\in G_{\geq0}$, which is equivalent to ensuring that the approximation for $L_t$ is in $\mathfrak{g}_{\geq 0}$, we need an additional assumption. For the Euler-Maruyama scheme to have results in $\mathfrak{g}_{\geq 0}$ it would be necessary that all increments $\Delta L_{t_k} \geq 0$, i.e. $L_t\geq 0$ must have monotonically increasing paths in time, as well. 

A class of processes satisfying this condition easily, would be all jump processes with positive jumps only.
Another possibility could involve processes with stochastic coefficients of the form
\begin{align*}
    dL_t^i &= a_i(t,Y_t^i) dt, \quad a_i(t,y)\geq 0\\
    dY_t^i &= b_i(t,Y_t^i)dt + c_i(t,Y_t^i) dS_t, \quad Y_0^i=y_0^i
\end{align*}
In this case, $L_t$ are positive, pathwise-increasing, continuous stochastic processes for any semimartingale $S_t$.

\begin{remark}\label{rem:decoupling}
Let us note, that decoupling the SDE in the Lie algebra does not mean that the SDE in the Lie group will be decoupled as well. On the contrary, one can see by the definition of the matrix exponential and the matrix multiplication therein that the resulting SDE will be fully coupled.

From a computational point of view, the decoupling in the Lie algebra is very advantageous, because all SDEs can be solved in parallel. Since we want to calibrate the SDE in the Lie group to historical rating matrices, it will be very important that the SDEs in the Lie algebra can be solved very fast.

From an analytical point of view, this approach translates the problem of defining an SDE with values in the space of stochastic matrices to simple SDEs taking values in $\R$, where a vast of literature and standard analytical tools are available.
\end{remark}

\section{Numerical tests}\label{sec:numerics}
In this section, we conduct two experiments, one for the direct exponential mapping and one for the geometric Euler approach. We calibrate the resulting rating models $\rSDE_t$ to $\rGAN_t$ at $t=1$, i.e. 1 year, by matching the first four moments. This is described in \Cref{sec:calibration} in more details. In \Cref{sec:directExp}, we show one example for the direct method using CIR processes on $\lieAlgebra_{\geq 0}$ and in \Cref{sec:gEM} we show another example for the geometric Euler approach using a constant drift and volatility.
In both sections, we will discuss the fit to the TimeGAN rating matrices by looking at their corresponding distributions at $1,3,6,12$ months and study some properties rating matrices should satisfy. These properties are introduced next in \Cref{sec:errors}. 



We used for the calibration of the rating SDE \matlab with the \matlabGOtoolbox
and for the training of the TimeGAN \python with\\ \tensorflow
running on \OS, on a machine with the following specifications: processor
\CPU and \RAM, and a \GPU.

\subsection{Rating properties}\label{sec:errors}
To estimate the quality of the TimeGAN and the SDEs we observed from the historical data that short term rating matrices up to one year should have the following properties: 
\begin{compactenum}
    \item It is more likely to stay in the initial rating than changing to another: This means rating matrices are strongly diagonal dominant, i.e.\ for $i=1,\dots,K$
        \begin{align}
            \left[R_{t}\left(\omega\right)\right]_{ii} \geq 
            \sum_{j \neq i}{\left[R_{t}\left(\omega\right)\right]_{ij}}.
            \label{eq:sDD}
        \end{align}
    \item Downgrading is more likely than upgrading: This means that the sum of the upper triangular matrix     is bigger than the sum of the lower triangular matrix, i.e.
        \begin{align}
            \sum_{i<j}{\left[R_{t}\left(\omega\right)\right]_{ij}}\geq 
            \sum_{i>j}{\left[R_{t}\left(\omega\right)\right]_{ij}}.
            \label{eq:dML}
        \end{align}
    \item Lower rated entities are more likely to default: This means that the default column is increasing from best starting rating to lowest, i.e. 
        \begin{align}
            \left[R_{t}\left(\omega\right)\right]_{1K}\leq 
            \left[R_{t}\left(\omega\right)\right]_{2K}\leq\dots\leq
            \left[R_{t}\left(\omega\right)\right]_{KK}.
            \label{eq:mDC}
        \end{align}
    \item The rating spreads more over time: We measure this by looking for decreasing diagonal elements, i.e.\ for all $s<t$ and all  $i=1,\dots,K$
        \begin{align}
            \left[R_{s}\left(\omega\right)\right]_{ii}\geq \left[R_{t}\left(\omega\right)\right]_{ii}.
            \label{eq:iRS}
        \end{align}
\end{compactenum}
These properties are not strict in the sense that they can be violated on some occasions. Moreover, one might think of other properties for rating matrices. Also for long term rating matrices (more than 1 year) these properties might not hold true anymore. This makes it very hard to define rigorous conditions for  rating matrices in general and are subject to future research and economical validation.

In \Cref{tab:trainingRP1} we can see a summary of the rating properties
\eqref{eq:sDD}--\eqref{eq:iRS} for the training data set. The numbers represent 
the percentages of time-sequences satisfying the conditions averaged over all initial ratings. For the rating spreads over time, we consider time steps from 0 to 1 month,
1 to 3, 3 to 6 and 6 to 12 and write down the percentages for $t=1,3,6,12$ respectively.
We can see that all of the rating matrices in the training data set were strongly diagonal dominant and nearly all had monotone increasing default columns.

The majority of the rating matrices put more emphasis on downgrading for time spans between one month and six months, while for one year all of them satisfied the condition. 

For the increasing rating spread we see the biggest violations of the property. This is most likely due to the fact that we consider all permutations of the data. It might be beneficial to filter these sequences out of the training set.

\begin{table}[htbp]
    \centering
    \caption{Rating properties for training data. Average percentage of the time series fulfilling the conditions \eqref{eq:sDD}--\eqref{eq:iRS}.}
    \begin{tabularx}{\linewidth}{X*{5}{C}}
    \rotateTableHead{Time in months} 
      & \rotateTableHead{Strongly diagonal dominant \eqref{eq:sDD}} 
      & \rotateTableHead{Downgrading is more likely \eqref{eq:dML}} 
      & \rotateTableHead{Monotone default column \eqref{eq:mDC}} 
      & \rotateTableHead{Increasing rating spread \eqref{eq:iRS}}\\
    \toprule
    1   & 100\,\%      & 87.96\,\%      & 100\,\%      & 100\,\%  \\      
    3   & 100\,\%      & 97.22\,\%      & 99.9\,\%     & 85.81\,\%  \\            
    6   & 100\,\%      & 94.44\,\%      & 100\,\%      & 83.18\,\%  \\    
    12  & 100\,\%      & 100\,\%        & 100\,\%      & 90.53\,\%  \\     
    \end{tabularx}
    \label{tab:trainingRP1}
\end{table}

In \Cref{tab:ganRP1} we see exactly the same table for TimeGAN using $M=12000$ synthetic time-sequences. Even though we did not impose any hard constraints, e.g.\ that rows must sum up to one, the DNN learned the conditions \eqref{eq:sDD}--\eqref{eq:iRS} very well,
as well as that rows must sum to one. The only criterion which was not always satisfied was again \eqref{eq:iRS} but less severe than for the training data.
Since these properties are almost always satisfied we did not optimize the hyperparameters or network architecture any further.

\begin{table}[htbp]
    \centering
    \caption{Rating properties for TimeGAN with $M=12000$. Average percentage of the time series fulfilling the conditions \eqref{eq:sDD}--\eqref{eq:iRS} and average row sums.}
    \begin{tabularx}{\linewidth}{X*{6}{C}}
    \rotateTableHead{Time in months} 
      & \rotateTableHead{Strongly diagonal dominant \eqref{eq:sDD}} 
      & \rotateTableHead{Downgrading is more likely \eqref{eq:dML}} 
      & \rotateTableHead{Monotone default column \eqref{eq:mDC}} 
      & \rotateTableHead{Increasing rating spread \eqref{eq:iRS}}
      & \rotateTableHead{Average row sums}\\
    \toprule
    1   & 100\,\%      & 100\,\%        & 100\,\%     & 100\,\%     & 0.9999  \\       
    3   & 100\,\%      & 100\,\%        & 100\,\%     & 100\,\%     & 0.9996  \\              
    6   & 100\,\%      & 100\,\%        & 100\,\%     & 93.2\,\%    & 1.0002  \\      
    12  & 100\,\%      & 100\,\%        & 100\,\%     & 93.33\,\%   & 1.0017  \\      
    \end{tabularx}
    \label{tab:ganRP1}
\end{table}

\subsection{Calibration of the rating SDE}\label{sec:calibration}
Before we start to explain, how we calibrate $\rSDE_t$ to $\rGAN_t$ let us explain why we do not calibrate directly to the historical data. Suppose that we select one specific time series of historical rating matrices and try to fit our model in a least-square sense in expectation. Then, the randomness should be eliminated by the optimizer since we want to fit all the different trajectories to one time sequence. This is not the way to go, if we desire a stochastic model for the rating transitions.
Another approach would be considering all of the training data set, sample as many trajectories and calibrate again in a least-square sense. There is no reason, why each of the random trajectories should match the particular rating matrix where it is subtracted from, maybe it would match another one perfectly.
So comparing trajectories does not make much sense either.

Hence, it makes more sense to compare distributions or moments of the data and the model. Now, the problem with using the historical rating matrices directly in this approach would be that at each specific point in time, we only have a few available matrices. Take for example the one year rating matrices, we only have 9 different matrices. Discussing a distribution of such a sample size is not very insightful.

Therefore, we rely on the ability of the TimeGAN to learn the behaviour of the time series of rating matrices. As aforementioned, considering the time series allows us to artificially inflate the data set by using all the permutations in time for the training. After the learning phase, we can sample fake time series data, getting an arbitrary number of different rating matrices at each point in time. Now, it makes sense to compare the moments of the fake rating matrices to the ones obtained at each point in time from $\rSDE_t$.

To be more precise, we use the standard estimators for mean, variance and moments of higher order in our experiments, i.e.\ for $k=3,\dots,n$, $n\in \N$,
\begin{align*}
    \left[\mu_{1}(t)\right]_{ij} &\coloneqq \frac{1}{M} \sum_{w=1}^{M}  \left[R_t(w)\right]_{ij},\\
    \left[\mu_{2}(t)\right]_{ij} &\coloneqq \frac{1}{M-1} \sum_{w=1}^{M}  \left(\left[R_t(w)\right]_{ij}-\left[\mu_{1}(t)\right]_{ij}\right)^2,\\
    \left[\mu_{k}(t)\right]_{ij}&\coloneqq \frac{1}{M} \sum_{w=1}^{M}  \left(\left[R_t(w)\right]_{ij}-\left[\mu_{1}(t)\right]_{ij}\right)^k.
\end{align*}
Let $\Pi$ denote the parameter set. Then, our objective function $f^n\colon\Pi \rightarrow \R^{n \cdot (K-1) \cdot K}$ is given by
\begin{align*}
    &f_k \colon\Pi \rightarrow \R^{(K-1) \cdot K},\quad
        f_k(p)\coloneqq \vectorize\left(\mu_{k}^{\mathrm{SDE}}(t;p) - \mu^\mathrm{GAN}_k(t)\right)\\
    &f^n(p)\coloneqq \left[w_1\cdot f_1(p),\dots,w_n\cdot f_n(p)\right]^T,
\end{align*}
where $w_k \in \R_{\geq 0}$ are weights and our minimisation problem can be formulated as a non-linear least square problem
\begin{align}
    \min_{p\in \Pi} \norm{f^n(p)}_2^2.
    \label{eq:nonLinearLeastSquare}
\end{align}
Of course, this procedure can be generalized by considering multiple points in time. Since this minimisation problem is very dependent on the performance of the DNN and its ability to learn the distribution of rating transition matrices from the historical data, one can also think of a penalized version of \eqref{eq:nonLinearLeastSquare}. For example one can add another least-square term for the most recent time series, i.e.
\begin{align*}
    \min_{p\in \Pi} \lambda_1 \norm{f^n(p)}_2^2 + 
    \lambda_2 \frac{1}{M}\sum_{w=1}^{M}\sum_{k=1}^{n}{\norm{\rSDE_{t_k}(w;p) - \rHistorical_{t_k}}_{F}^{2}},
\end{align*}
where $\norm{\cdot}_F$ denotes the Frobenius norm and $\lambda_1, \lambda_2 \geq 0$ are weights. We will make the code publicly available and leave this experiment for the reader.

\begin{remark}\label{rem:MoreRatings}
As aforementioned, using rating matrices with more than four ratings is straightforward in this approach. 
Since the SDEs in the Lie algebra are decoupled and can be computed in parallel, solving them will not lead to a major performance bottleneck compared to fewer ratings. The more relevant issue is that the number of parameters in the calibration increases quadratically, making it more and more important to use some principle component analysis to make the calibration more efficient. A possibility to use the autoencoder of the TimeGAN comes to mind, this is however subject to future research.

Also it is straightforward to remove the condition that the default rating is absorbing. In this case, we would need $(K-1)\cdot K$ decoupled SDEs in the Lie algebra.
\end{remark}

\subsection{The case of direct exponential mapping.}\label{sec:directExp} 
Let us now consider $\rCIR_t\coloneqq \exp\left(L_t\right)$, where
\begin{align*}
    dL_t^i = a_i \left(b_i -L_t^i \right)dt + \sigma_i \sqrt{L_t^i} dW_t.
\end{align*}
Each of the SDEs have a parameter for the mean-reversion $b_i$, mean-reversion speed $a_i$ and volatility $\sigma_i$, which are all assumed to be positive. During our calibration procedure we allow the Feller-condition to be violated for simplicity.
The parameter set is therefore given by positive real numbers 
$\Pi^{\mathrm{CIR}}\coloneqq \R^{3\cdot (K-1)^2}_{\geq 0}$ by stacking the individual parameters below each other.
We found during our experiments that values between zero and one worked best. 
We calibrated $\rCIR_t$ for $t=1$, i.e. for the 12 month rating transitions, by matching the moments up to order 4. For the variance we added a weight $w_2=10$ and set $w_1=w_3=w_4=1$ to put more emphasis on the variance. The corresponding parameters after the calibration procedure with $M=1000$ trajectories for $\rCIR_t$ and $M=10000$ trajectories for $\rGAN_t$ can be found in \Cref{tab:cirParamCal1}. The first column explains to which basis element the coefficients belong. To be more precise, 2-3 means that the initial rating is 2 and at $t=1$ we transition to rating 3. The minimisation error \eqref{eq:nonLinearLeastSquare} in this case was
$7.494e-05$, telling us that the moments up to order 4 match very well and it took roughly 68.5 seconds using \lsqnonlin with the Trust-Region-Reflective algorithm.

\begin{figure}
    \centering
    \includegraphics[width=\columnwidth]{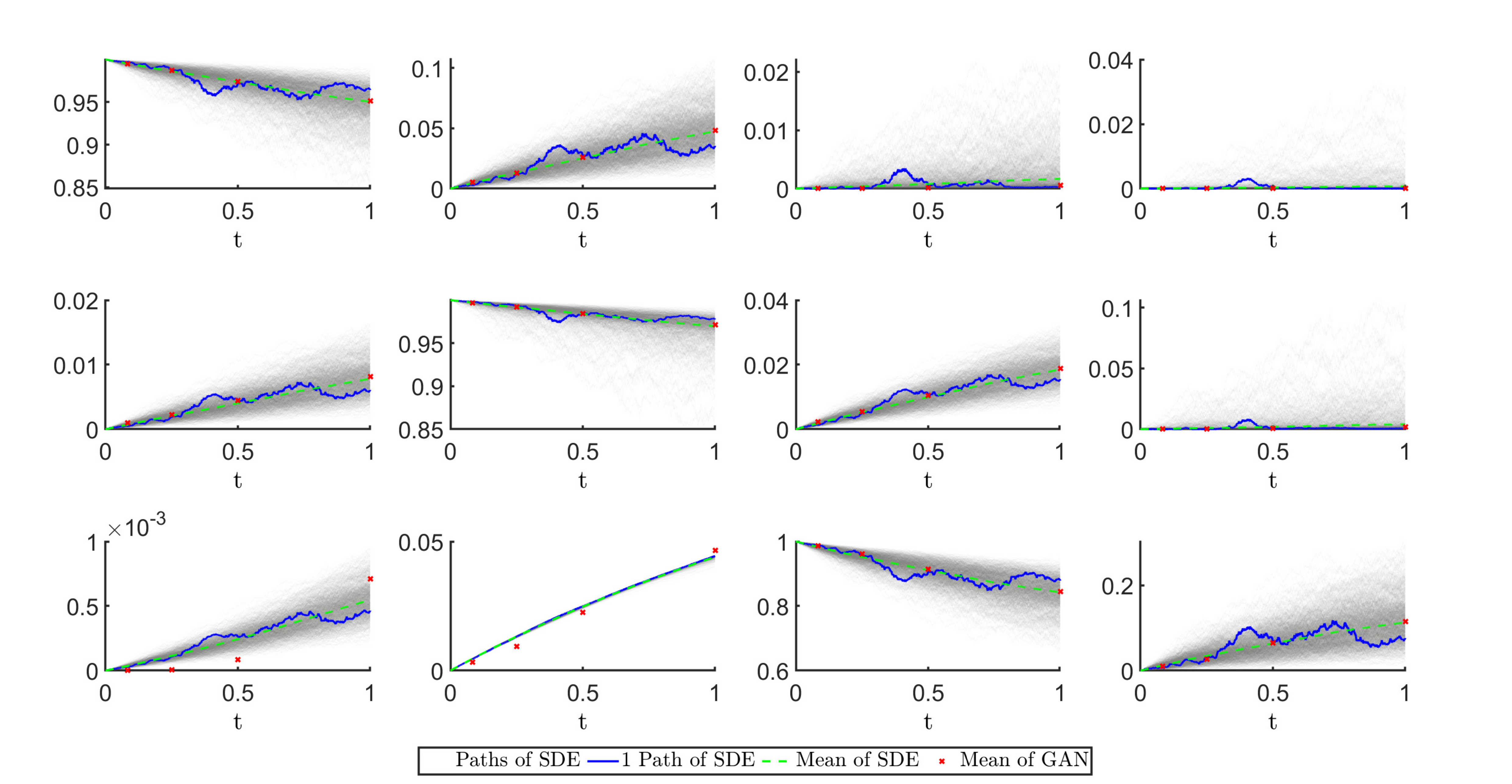}
    \caption{Trajectories of calibrated $\rCIR_t$ with parameters as in \Cref{tab:cirParamCal1}.}
    \label{fig:cirTra1}
\end{figure}

\begin{table}[htbp]
    \centering
    \caption{Parameters of $\rCIR$ after calibration at $t=1$ to $\rGAN$ using $n=4$ moments.}
    \begin{tabular}{*{4}{c}}
        From-To & $a$ & $b$ & $\sigma$\\
        \toprule
        1-2 & 2.41e-01 & 2.29e-01 & 1.28e-01\\
        1-3 & 3.73e-02 & 3.73e-02 & 1.17e-01\\
        1-4 & 6.80e-02 & 6.74e-03 & 1.21e-01\\
        2-1 & 9.25e-02 & 9.19e-02 & 4.59e-02\\
        2-3 & 1.50e-01 & 1.47e-01 & 6.01e-02\\
        2-4 & 5.34e-02 & 5.44e-02 & 2.47e-01\\
        3-1 & 2.06e-02 & 2.01e-02 & 6.74e-03\\
        3-2 & 3.01e-01 & 1.87e-01 & 9.14e-03\\
        3-4 & 4.07e-01 & 3.69e-01 & 2.62e-01
    \end{tabular}
    \label{tab:cirParamCal1}
\end{table}
In \Cref{fig:cirTra1}, we can see the trajectories of $\rCIR_t$ over time for each entry in the rating matrix except for the last row. The upper left corner are the transition probabilities from \rFormat{A}
to \rFormat{A}, right next to it from \rFormat{A} to \rFormat{B} and so on. The grey lines are a cloud of 
$1000$ trajectories of $\rCIR_t$ and the blue line is one trajectory. The green dashed line is the mean at each time of the process and the red dots are the means of $\rGAN_t$ at $t=1,3,6,12$ months.
We can see that the paths are rough and the mean-reversion of the CIR processes is apparent as well, since the blue line tends to come back to the green dashed line illustrating its mean. Also we see again a good fit over time to $\rGAN_t$ by comparing how close the mean of $\rCIR_t$ is compared to the mean of $\rGAN_t$.

\begin{remark}\label{rem:SDEonLieGroup}
    We modelled the rating transition by starting with an SDE on the positive half-space of the
    Lie algebra of stochastic matrices. Another approach could involve, modelling the SDE on 
    the appropriate half-space of the Lie group directly. To do this, it would be necessary 
    to use SDEs respecting the underlying geometry, i.e.\ Stratonovich-SDEs, since they obey the
    chain rule, or the Itô counterpart by Itô-Stratonovich conversion.
    
    In this line of research, numerical methods such as Runge-Kutta-Munthe-Kaas (RKMK) or the Magnus expansion are available, see for instance \cite{Muniz2022} and \cite{KPP2021} for more details.
    
    The advantage of studying these SDEs directly on the Lie group are that one can check more easily if the SDE will satisfy the rating matrix properties.
\end{remark}

\begin{figure}[h]
    \centering
    \includegraphics[width=\columnwidth]{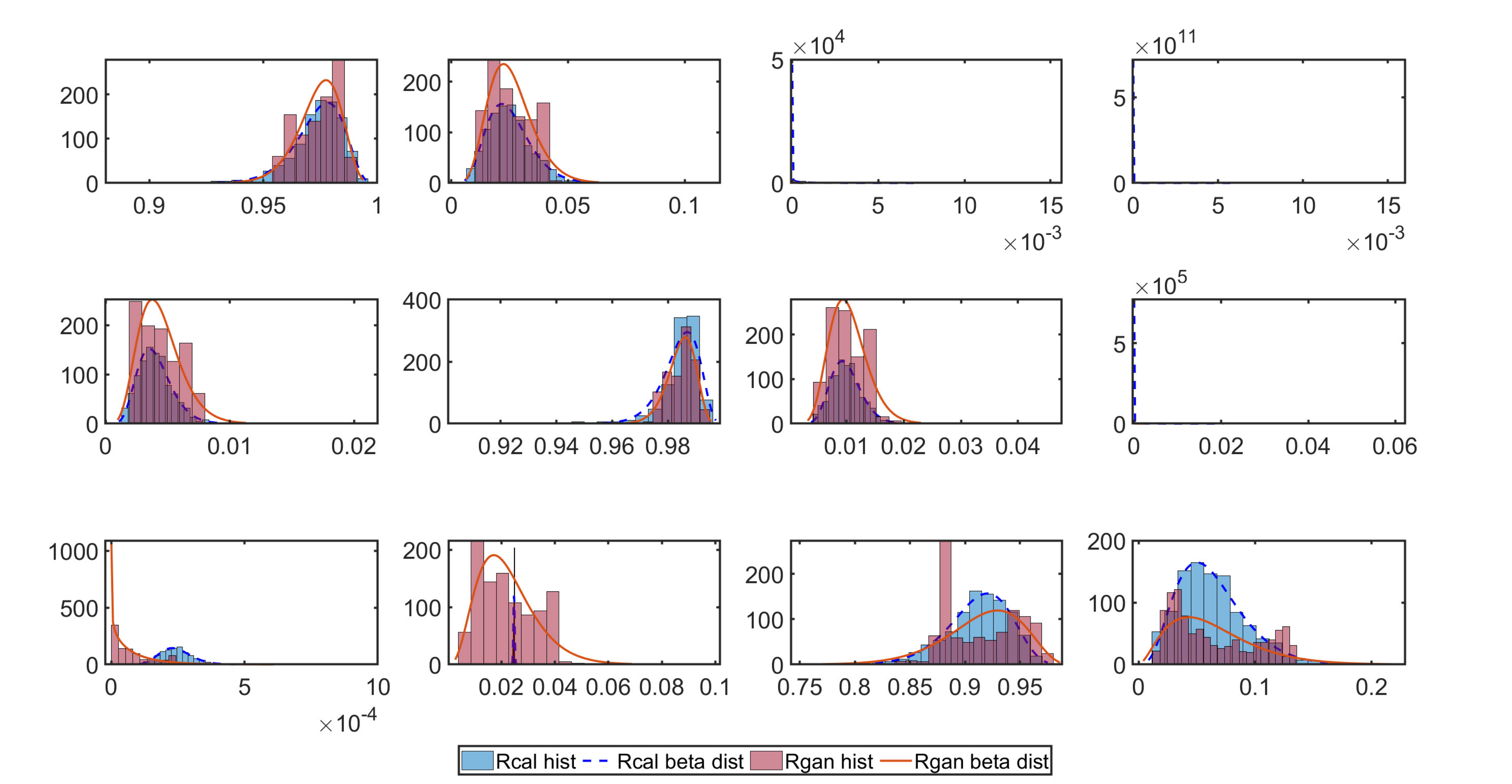}
    \caption{Histograms of ratings transition probabilities at 6 months.}
    \label{fig:cirRD3}
\end{figure}

\begin{figure}[h]
    \centering
    \includegraphics[width=\columnwidth]{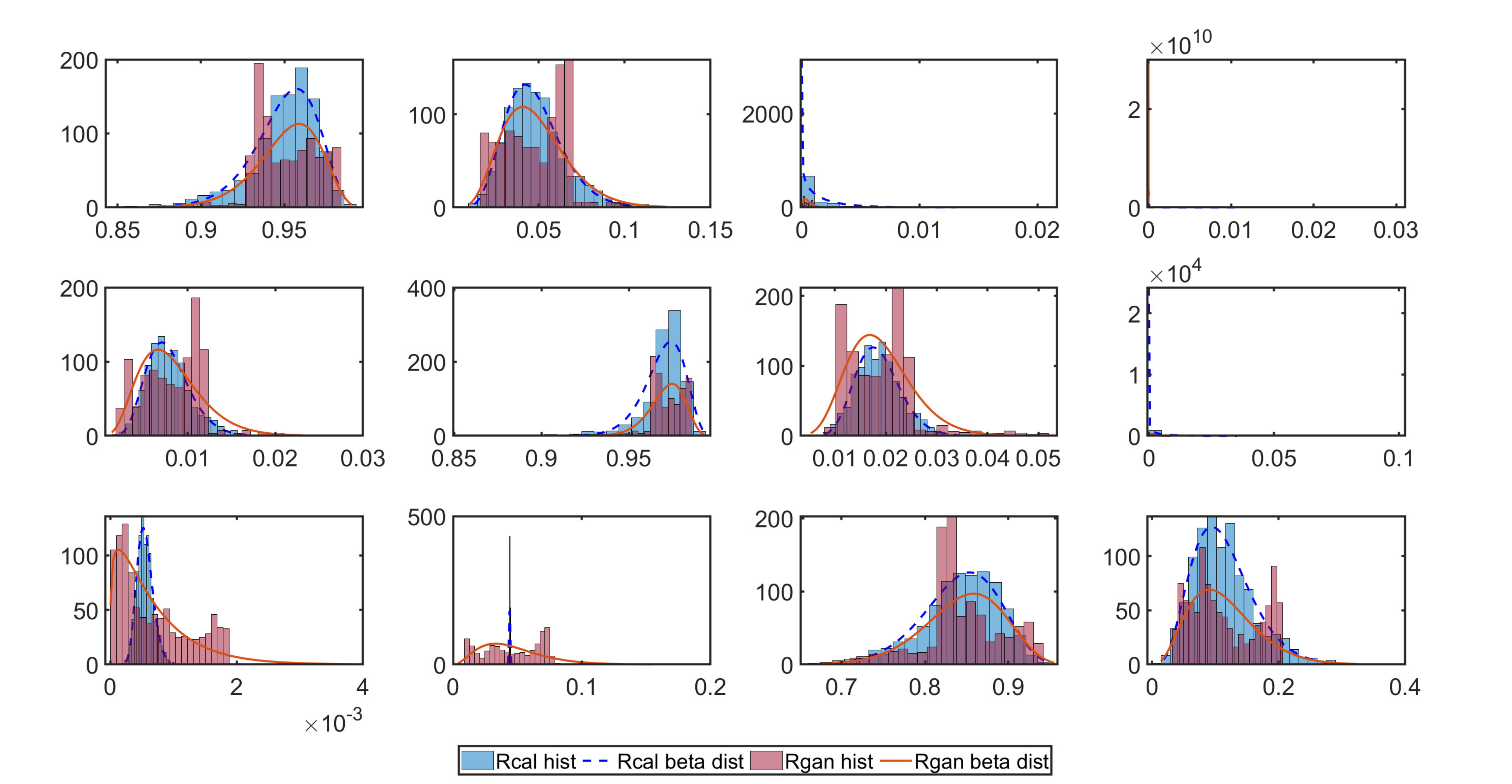}
    \caption{Histograms of ratings transition probabilities at 12 months.}
    \label{fig:cirRD4}
\end{figure}

\paragraph*{Analysis of the rating distributions and properties}
Since we expect that downgrades are more likely than upgrades, we expect that the rating distributions should be skewed with one tail being fatter than the other. We can see this in both \Cref{fig:cirRD3} ($t=0.5$) and \Cref{fig:cirRD4} ($t=1$). Each of the figures are ordered as the entries for the rating matrices excluding the last row. This means that the upper left subfigure shows the transitions for \rFormat{A} to \rFormat{A}, the one right next to it  
\rFormat{A} to \rFormat{B} and so on. The red columns are the histogram of $\rGAN_t$ and the blue columns illustrate the histogram of $\rCIR_t$. We fitted beta distributions to the histograms. The red solid line is the according beta distribution of $\rGAN_t$ and
the blue dashed line the beta distribution of $\rCIR_t$.

Let us focus for the moment on \Cref{fig:cirRD4}, i.e. the rating transitions for one year.
The distributions using $\rGAN$ look like they have two modes and suggest a mixture Gaussian model.
Therefore, the beta distributions do not describe the data very well. However, we have no intuition why the rating transitions should have two modes and consider it as subject for further investigation.

For $\rCIR_t$ we see a close match of the beta distribution to the histograms and match our initial intuition that the model should have one tail being fatter than the other.

In \Cref{fig:cirRD3} we see in most of the subfigures a good match of the shapes of the beta distributions of $\rGAN$ and $\rCIR$ even though the CIR processes have constant coefficients and are calibrated to the moments of $\rGAN$ at $t=1$. We saw the same for $t=1,3$ months and therefore decided not to put the figures to shorten the presentation.

Let us now assess the quality of the model rating matrices as for the training data set and TimeGAN
by \eqref{eq:sDD}--\eqref{eq:iRS}. \Cref{tab:cirRP1} is structured exactly like \Cref{tab:trainingRP1}. We can see similar results to \Cref{tab:ganRP1}. Almost all the conditions are satisfied perfectly except for \eqref{eq:iRS}, where only $7\,\%$ violated the condition at $t=6,12$ months. Another downside of this method can be seen in \Cref{fig:cirTra1} by focusing on the blue trajectory in the default-column.
It seems possible that the default is not absorbing because the trajectories are not monotonically increasing, only the mean is increasing. This could be viable if we allow companies to recover from default over time if they were not bankrupt from begin with, which in fact would be more realistic, because otherwise either every entity would eventually default or at some point no entity would default anymore. Also it could be interesting to study conditions in this setting to ensure monotone increasing paths in the default column, which is subject to future research. We will see in the next section that the geometric Euler approach will not suffer from this problem.
\begin{table}[htbp]
    \centering
    \caption{Rating properties for $\rCIR$. Average percentage of the time series fulfilling the conditions \eqref{eq:sDD}--\eqref{eq:iRS}.}
    \begin{tabularx}{\linewidth}{X*{5}{C}}
    \rotateTableHead{Time in months} 
      & \rotateTableHead{Strongly diagonal dominant \eqref{eq:sDD}} 
      & \rotateTableHead{Downgrading is more likely \eqref{eq:dML}} 
      & \rotateTableHead{Monotone default column \eqref{eq:mDC}} 
      & \rotateTableHead{Increasing rating spread \eqref{eq:iRS}}\\
    \toprule
    1   & 100\,\%      & 100\,\%    & 100\,\%      & 100\,\%  \\      
    3   & 100\,\%      & 100\,\%    & 100\,\%      & 100\,\%  \\            
    6   & 100\,\%      & 100\,\%    & 100\,\%      & 93.22\,\%  \\    
    12  & 100\,\%      & 100\,\%    & 100\,\%      & 93.34\,\%  \\     
    \end{tabularx}
    \label{tab:cirRP1}
\end{table}
\subsection{The case of geometric Euler Maruyama.}\label{sec:gEM}
Let us now consider $\rGEM_t$ and assume that each of the SDEs are given by 
\begin{align*}
    dL_t^i &= \abs{Y_t^i}^{a_i} dt\\
    dY_t^i &= b_i dt + \sigma_i dW_t,\quad Y_0^i=0.
\end{align*}
They have a parameter for a constant drift $b_i$, power $a_i$ and volatility $\sigma_i$, which are all assumed to be positive.
The parameter set is therefore given by positive real numbers 
$\Pi^{\mathrm{gEM}}\coloneqq \R^{3\cdot (K-1)^2}_{\geq 0}$ by stacking the individual parameters below each other.
We found during our experiments that values between zero and two worked best. 
We calibrated $\rGEM_t$ for $t=1$, i.e. for the 12 month rating transitions, by matching the moments up to order 4. For the variance we added a weight $w_2=10$ and set $w_1=w_3=w_4=1$ to put more emphasis on the variance. The corresponding parameters after the calibration procedure with $M=1000$ trajectories for $\rGEM_t$ and $M=10000$ trajectories for $\rGAN_t$ can be found in \Cref{tab:gemParamCal1}. The first column explains to which basis element the coefficients belong. To be more precise, $2-3$ means starting rating is 2 and at $t=1$ we transition to rating 3. The minimisation error \eqref{eq:nonLinearLeastSquare} in this case was
$5.265e-05.$, telling us that the moments up to order 4 match very well and it took roughly 1058 seconds using \lsqnonlin with the Trust-Region-Reflective algorithm.

\begin{figure}[h]
    \centering
    \includegraphics[width=\columnwidth]{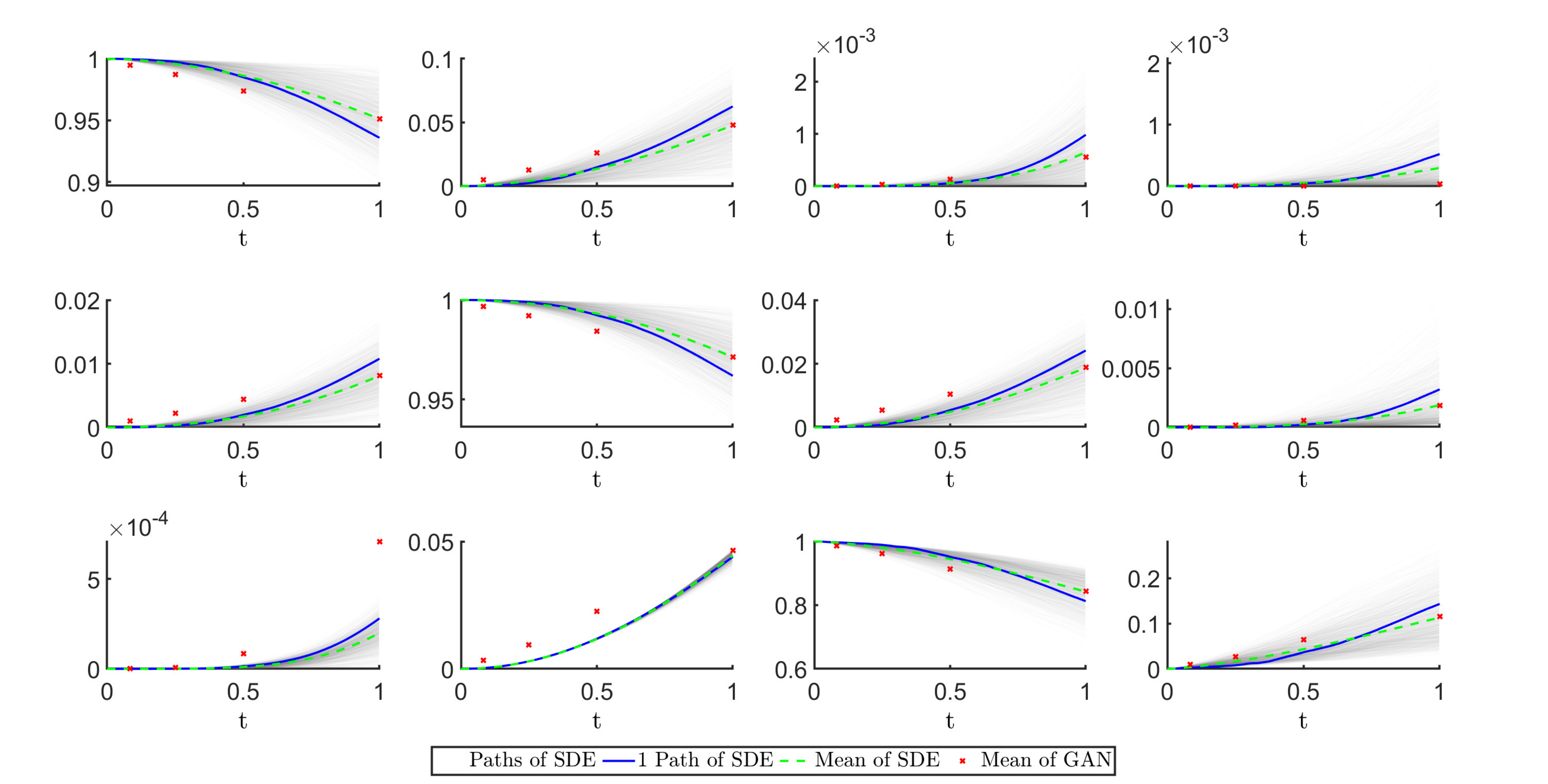}
    \caption{Trajectories of calibrated $\rGEM_t$ with parameters as in \Cref{tab:gemParamCal1}.}
    \label{fig:gemTra1}
\end{figure}

\begin{figure}[h]
    \centering
    \includegraphics[width=\columnwidth]{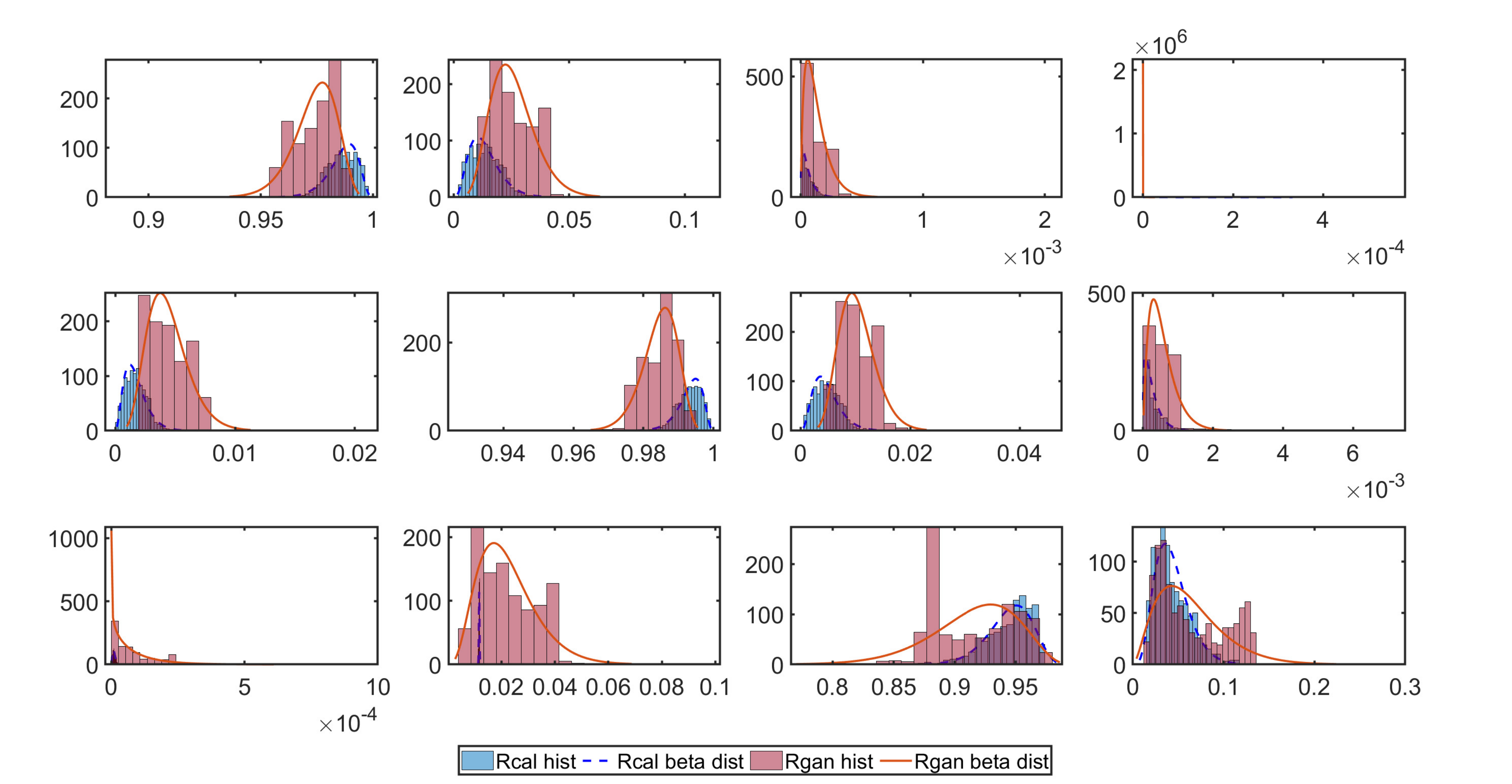}
    \caption{Histograms of ratings transition probabilities at 6 months.}
    \label{fig:gemRD3}
\end{figure}

\begin{figure}[h]
    \centering
    \includegraphics[width=\columnwidth]{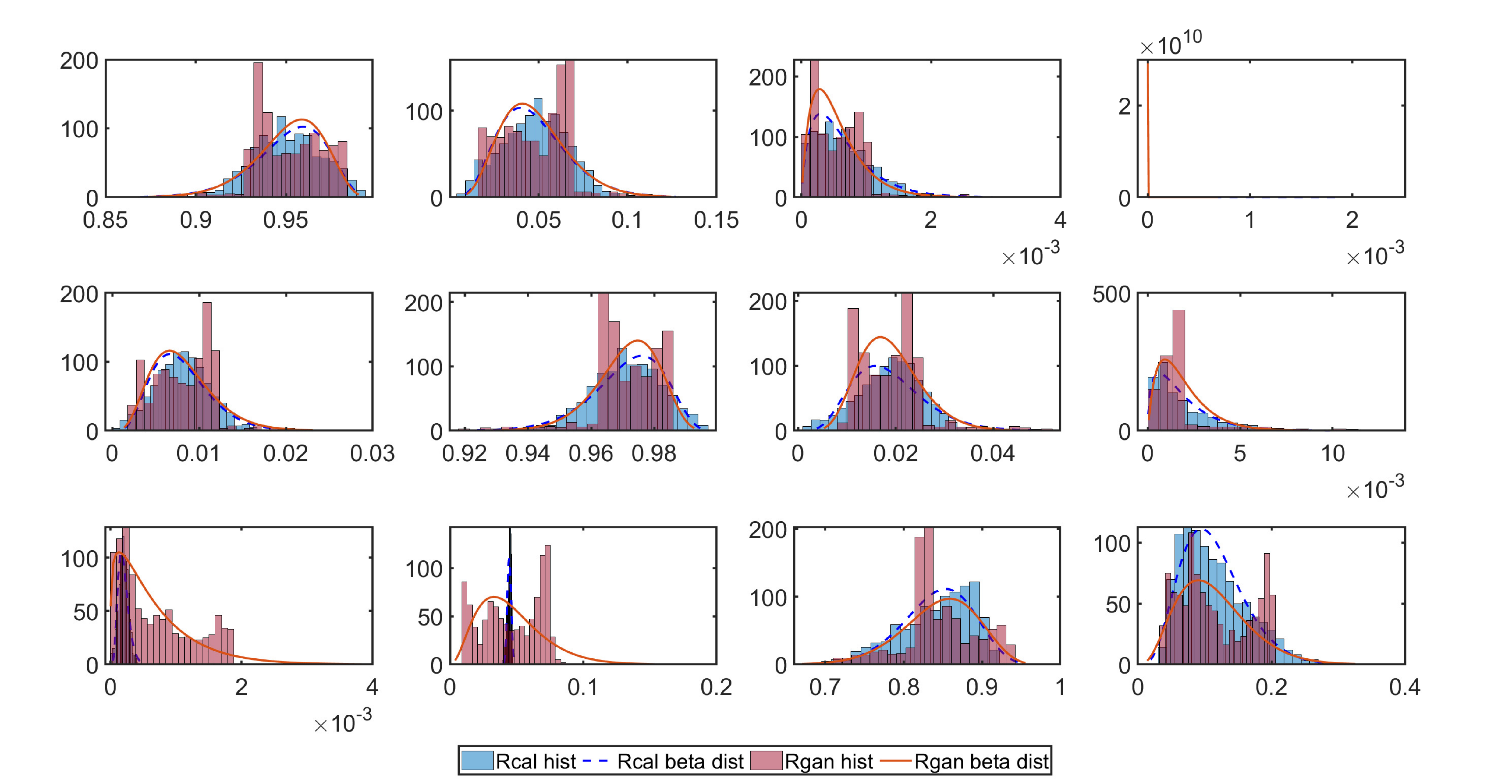}
    \caption{Histograms of ratings transition probabilities at 12 months.}
    \label{fig:gemRD4}
\end{figure}

\begin{table}[htbp]
    \centering
    \caption{Parameters of $\rGEM$ after calibration at $t=1$ to $\rGAN$ using $n=4$ moments.}
    \begin{tabular}{*{4}{c}}
        From-To & $a$ & $b$ & $\sigma$\\
        \toprule
        1-2 & 9.21e-01 & 7.70e-02 & 3.15e-02\\
        1-3 & 1.85e+00 & 9.61e-03 & 5.57e-03\\
        1-4 & 1.92e+00 & 1.41e-02 & 1.50e-02\\
        2-1 & 1.32e+00 & 4.91e-02 & 1.39e-02\\
        2-3 & 1.09e+00 & 5.43e-02 & 1.74e-02\\
        2-4 & 1.86e+00 & 2.13e-02 & 2.58e-02\\
        3-1 & 1.99e+00 & 8.17e-04 & 1.00e-04\\
        3-2 & 1.03e+00 & 1.09e-01 & 1.00e-04\\
        3-4 & 8.03e-01 & 7.59e-02 & 1.38e-01
    \end{tabular}
    \label{tab:gemParamCal1}
\end{table}
In \Cref{fig:gemTra1} we can see the trajectories of $\rGEM_t$ over time for each entry in the rating matrix except for the last row. The upper left corner are the transition probabilities from \rFormat{A}
to \rFormat{A}, right next to it from \rFormat{A} to \rFormat{B} and so on. The grey lines are a cloud of 
$1000$ trajectories of $\rGEM_t$ and the blue line is one trajectory. The green dashed line is the mean at each time of the process and the red dots are the means of $\rGAN_t$ at $t=1,3,6,12$ months.
We can see that the paths are much smoother compared to \Cref{fig:cirTra1}. Also we see again a good fit at the terminal time to $\rGAN_t$ by comparing how close the mean of $\rGEM_t$ is compared to the mean of $\rGAN_t$. For $t=1,3,6$ months we see a slight deviation of their corresponding means, suggesting that we should either use time-dependent parameters or different SDEs.
\paragraph*{Analysis of the rating distributions and properties}
In \Cref{fig:gemRD3} and \Cref{fig:gemRD4} we can see the analogue of 
\Cref{fig:cirRD3} and \Cref{fig:cirRD4} from \Cref{sec:directExp}. We used the same trajectories of $\rGAN_t$ in these plots to be able to compare both methods amongst each other.

Let us focus for the moment on \Cref{fig:gemRD4}, i.e.\ the rating transitions for one year.
For $\rGEM_t$ we see a close match of the beta distribution to the histograms as well. Also we see a very good fit of the beta distributions of $\rGAN$ and $\rGEM$ towards each other. This fit looks even closer than in \Cref{fig:cirRD4} for $\rGAN$ and $\rCIR$.

In \Cref{fig:gemRD3}, the six month rating transitions, we see a worse fit to the data than we saw in \Cref{fig:cirRD3} using $\rCIR_t$.
This suggests that one should either use a different underlying SDE for $Y_t^i$ or introduce time-dependent parameters.
Most remarkably all the conditions \eqref{eq:sDD}--\eqref{eq:iRS} were satisfied perfectly for this model.
\section{Conclusion and future research}\label{sec:Conclusion}
In this paper, we developed a novel methodology in the community of rating transition modelling, making it possible
to formulate rating transitions as processes on Lie groups by using its relation to its Lie algebra and imposing SDEs there. We showed two different approaches, first the direct exponential mapping in \Cref{sec:directExp} and showed numerical results using CIR processes in the Lie algebra.
Second we demonstrated, how the geometric Euler method can be applied to preserve the Chapman-Kolmogorov equations in \Cref{sec:gEM}. In \Cref{tab:conclusion1} we compare the two methods and their features.

\begin{table}[h]
    \centering
    \begin{tabularx}{\linewidth}{*{1}{X}*{1}{X}}
        \multicolumn{1}{c}{$\rCIR_t$} & \multicolumn{1}{c}{$\rGEM_t$}\\
        \toprule
        Simple method with fast calibration & More complex with slower calibration\\
        Needs only $L_t^i$ to be positive & Requires that $L_t^i$ has monotonically increasing paths\\ 
        Satisfies all rating properties well & Satisfies all rating properties perfectly\\
        Does not satisfy the Chapman-Kolmogorov equations & Satisfies Chapman-Kolmogorov equations\\
        Default column is not absorbing & Default column is absorbing\\
    \end{tabularx}%

    \caption{Comparison of $\rCIR_t$ and $\rGEM_t$.}
    \label{tab:conclusion1}
\end{table}

As mentioned at various points throughout this paper, there are many possibilities for future research.

For instance, we could try to learn the historical generators instead of the rating transitions. In this case, we would be able to calibrate the SDE on the Lie algebra to the fake generators. Also novel neural network architectures called DeepONets (cf. \cite{Lu2019}) could be thought of in this framework.

Another line of research could involve adding an additional network to the TimeGAN which outputs the calibrated parameters of the target SDE directly. It would be beneficial to link the Autoencoder or Supervisor network to this new network to exploit dimensionality reductions.

In a next step, we would like to include the possibility to furthermore calibrate the rating SDE to Credit-Default-Swap (CDS) quotes
under the risk-neutral measure. This extension will be useful for instance in the context of rating triggers under a netting agreement with Credit-Support-Annex (CSA) for valuation adjustments.
\appendix
\section*{Declarations}
\subsection*{Funding}
This project has received funding from the European Union’s Horizon 2020 research and innovation
programme under the Marie Sklodowska-Curie grant agreement No 813261 and is part of the ABC-EU-XVA project.
\subsection*{Conflicts of interests}

The authors have no relevant financial or non-financial interests to disclose.

\subsection*{Data availability}
All data generated or analysed during this study are included in this published article except 
for the historical rating transition data, which has to be downloaded from the respective websites
of the rating agencies while agreeing to their terms of usage.
The code and data sets to produce the numerical experiments are available at
\url{https://github.com/kevinkamm/RatingML}.
{\thispagestyle{scrheadings}
\thispagestyle{scrheadings}\ihead{}
\singlespacing
\begin{footnotesize}
\bibliographystyle{acm}
\bibliography{literature.bib}
\end{footnotesize}
}
\end{document}